\documentclass[onecolumn,a4paper,IEEEtrans]{article}
\usepackage[margin=0.75in]{geometry}
\usepackage{amsmath,graphicx}

\usepackage{subcaption}
\usepackage{mwe}
\usepackage{mathtools, nccmath} 

\usepackage{blindtext}
\usepackage{hyperref}
\usepackage{listings}
\usepackage{lipsum}
\newsavebox{\codebox}

\usepackage{tabularx}
\usepackage[table]{xcolor}
\usepackage{colortbl}
\definecolor{lightgreen}{RGB}{200,255,200}
\definecolor{lightred}{RGB}{210,100,100}
\definecolor{lightblue}{RGB}{200,220,255}

\newcommand{\gcell}{\cellcolor{lightgreen}}
\newcommand{\bcell}{\cellcolor{lightblue}}

\def\code#1{\texttt{#1}}

\title{SMRVIS: Point cloud extraction from 3-D ultrasound \\for non-destructive testing}
\date{June 5, 2023}
\author{Lisa Y.W. Tang}
 
\begin{document}
\maketitle


\begin{abstract}

We propose to formulate point cloud extraction from ultrasound volumes as an image segmentation problem. Through this convenient formulation, a quick prototype exploring various variants of the Residual Network, U-Net, and the Squeeze and Excitation Network was developed and evaluated. This report  documents the experimental results compiled using a training dataset of five labeled ultrasound volumes and 84 unlabeled volumes that got completed in a two-week period as part of a submission to the open challenge ``3D Surface Mesh Estimation for CVPR workshop on Deep Learning in Ultrasound Image Analysis''. Based on external evaluation performed by the challenge's organizers, the framework came first place on the challenge's \href{https://www.cvpr2023-dl-ultrasound.com/}{Leaderboard}. Source code is shared with the research community at  a \href{https://github.com/lisatwyw/smrvis}{public repository}.
\end{abstract}
\vspace{1em}
\emph{Keywords:
Ultrasound volumes; Non-Destructive Testing; Pipe; Manufacture Defects; Attention U-Net; Recurrent-Residual U-Net;  Squeeze-Excitation U-Net; U-Net++; W-Net}

\DeclarePairedDelimiter{\nint}\lfloor\rceil
\DeclarePairedDelimiter{\abs}\lvert\rvert 

\section{Introduction}
\label{sec:intro}





As part of a submission to an open challenge entitled ``3D Surface Mesh Estimation for Computer Vision and Pattern Recognition workshop on Deep Learning in Ultrasound Image Analysis'', this report presents experimental results compiled using a training dataset of five labeled ultrasound volumes and 85 unlabeled volumes. Due to resource and time constraints \cite{winner23}, we focus on finding a workable solution that can be quickly prototyped and tested. Accordingly, we propose to formulate three-dimensional mesh estimation from ultrasound volumes as an image segmentation problem. Source code is shared with the research community at 
\url{https://github.com/lisatwyw/smrvis}. 
We nicknamed this proposed framework as SMRVIS (Surface Mesh Reconstruction Via Image Segmentation) partly to draw an analogy to an older framework called SMRFI \cite{smrfi} that formulated shape matching as a feature image registration problem. Unlike this prior work that embedded 2D shapes with feature values and assigned the computed feature values to the nearest voxel of the embedding space, the present framework simply encodes the positions of the reference mesh vertices with values in the range of [0, 1].
\\\\
Rather than performing segmentation in 3D, our current approach performs a series of segmentations of thin sections by treating overlapping consecutive slices as 3-channel inputs. Our mesh-embedding scheme does not limit us to the deployment of two-dimensional models only. However, as the number of labeled data in this challenge dataset is relatively small ($n=$5), the choice of two-dimensional models allowed us to adopt a patch-based training approach, which has countless success cases as observed in previous open challenges \cite{winner23}. To this end, we have explored and evaluated W-Net, R2 U-Net, SE U-Net, Attention U-Net, and U-Net++, which are variants of the U-Net \cite{wnet}. 



\subsection{Background}

Nowadays, a vast majority of segmentation frameworks employ an encoder-decoder neural network structure that is popularized by the U-Net architecture \cite{wnet}. In this architecture, network components called \textbf{skip connections} allow data from down-sampling layers to be rerouted back to the up-sampling layers. 
\\\\
Since its initial success, researchers have proposed countless variants of U-Net to enhance its performance \cite{nnunet,heinrich23}. For instance,  Xia et al. developed the W-Net in 2017 that joins two fully convolutional neural network (CNN) branches with an autoencoder such that the first branch would encode input data into a fuzzy segmentation while the second branch would reconstruct the fuzzy segmentation to input data. Since its proposal, the W-Net is shown to have numerous successful applications such as retinal vessel segmentation \cite{retinal} and generation of Chinese characters \cite{chinese}.  
\\\\
In 2017, Hu et al. further enhanced U-Nets by incorporating the \textbf{squeeze and excitation blocks} at the end of each convolutional block in order to enhance the inter-dependencies between channels. In 2018, Oktay et al. proposed to incorporate an \textbf{attention module} within each skip connection that will drive the overall model to focus on input regions that garner more importance. Around the same time, Zhou et al. proposed the U-Net++ that aggregates features across different scales via its specially designed skip connections while Alom et al. \cite{r2unet} proposed the recurrent residual (R2Unet) U-Net, which combines the strengths of recurrent connections and residual networks and was shown to have improved the quality of the feature representations they could produce \cite{r2unet}. 
\\\\
In 2022, Kugelman et al. \cite{Kugelman} conducted a comparative study to benchmark some of the aforementioned U-Net variants in the context of retinal tissue extraction from optical coherence tomography and recommended the adoption of R2-U-Nets. 
\\\\
We adopted in this work their opensource implementation \cite{Kugelman} that employs the same \textbf{deconvolutional block} for all U-Net variants. Code listing \ref{lst:unet} provides an abstraction of a basic U-Net implementation. An accompanying notebook is available at \url{https://github.com/lisatwyw/unet_variants/blob/main/tf_U_Net.ipynb}. 

\begin{lrbox}{\codebox}
\begin{lstlisting}[label={lst:unet},caption={Schematics of basic U-Net},language=python,basicstyle=\small]
from tensorflow.keras.layers import BatchNormalization, Add, Multiply, Concatenate
from tensorflow.keras.layers import Input, ConvND, ConvNDTranspose 
from tensorflow.keras.layers import GlobalAveragePoolingND, MaxPoolingND 
from tensorflow.keras.models import Model

# above function names are abstractions only (ND instead of 2D or 3D)

def stack_bn_act( x, NF, KS  ):          
  for i, ks in enumerate( KS ):    
    x = Conv2D( NF, ks, padding='same' )(x)
    x = BatchNormalization()(x)
    x = Activation('relu')(x)        
  return x

def conv_block( x, NF, KS ):  
  o = Conv2D( NF, kernel_size= KS[0], padding='same' )(x)
  o = BatchNormalization()(o)
  o = Activation( 'relu' )(o)    
  o = stack_bn_act(o, NF, KS[1:]) # defined above   
  return o

def deconv_block( x, NF, KS=2 ):    
  o = Conv2DTranspose( NF, KS, padding='same' )(x)
  o = BatchNormalization()(o)
  o = Activation( 'relu' )(o)                      
  return o
  
# -------------- hyperparameters tested in ablation study
AC, BS, NF = 'sigmoid', 8, 16
NX, NY NDIM, n_slices=  224, 224, 2, 3
ks=[(1,1)]*len(nfilters)

inp = Input( (NX, NY, 3) )

o1 = conv_block( inp, NF,    ks_s ); p1 = MaxPooling2D()( o1 )                
o2 = conv_block( p1,  NF*2,  ks_s ); p2 = MaxPooling2D()( o2 )                
o3 = conv_block( p2,  NF*4,  ks_s ); p3 = MaxPooling2D()( o3 )                
o4 = conv_block( p3,  NF*8,  ks_s ); p4 = MaxPooling2D()( o4 )                
o5 = conv_block( p4,  NF*16, ks_s ); p5 = MaxPooling2D()( o5 )                

o6 = conv_block( Concatenate()( [deconv_block( o5, NF*8, strides=(2,2) ), o4]), NF*8, ks)
o7 = conv_block( Concatenate()( [deconv_block( o6, NF*4, strides=(2,2) ), o3]), NF*4, ks)
o8 = conv_block( Concatenate()( [deconv_block( o7, NF*2, strides=(2,2) ), o2]), NF*2, ks)
o9 = conv_block( Concatenate()( [deconv_block( o8, NF*1, strides=(2,2) ), o1]), NF*1, ks)

out = Activation( AC )( Conv2D( n_slices, 1 )( o9 ) )
model = Model(inp, outputs=out )
\end{lstlisting}
\end{lrbox}

\noindent 
\colorbox{gray!10}{
\setlength{\fboxrule}{0ex} 
\fbox{%
  \begin{minipage}{\dimexpr\linewidth-0ex}
  \usebox{\codebox}    
  \end{minipage}%
  }
}

\begin{lrbox}{\codebox}
\begin{lstlisting}[label={lst:att},caption={Select network components highlighted in the main text of Section 1.1},language=python,basicstyle=\small]

def attention_block( input_block, gate, ks=(1,1) ):    
  x = Conv2D( NF, ks )(input_block)
  x = BatchNormalization()(x)

  g = Conv2D( NF, ks )(gate)
  g = BatchNormalization()(g)    
  
  att_map = Add()( [g, x] )    
  att_map = Activation( 'relu' )(att_map)
  
  att_map = Conv2D( 1, ks )(att_map)
  att_map = Activation( 'sigmoid')(att_map)
  x = Multiply()( [input_block, att_map ] ) 

  return x

... # identical to Code Listing 1         
o5 = conv_block( p4,  NF*16, ks_s ); p5 = MaxPooling2D()( o5 )                

c6=attention_block( deconv_block( o5, NF*8, strides=(2,2) ), o4) 
c7=attention_block( deconv_block( o6, NF*4, strides=(2,2) ), o3) 
c8=attention_block( deconv_block( o7, NF*2, strides=(2,2) ), o2) 
c9=attention_block( deconv_block( o8, NF*1, strides=(2,2) ), o1) 

o6 = conv_block( Concatenate()( [c6, o4] ), NF*8, ks )
o7 = conv_block( Concatenate()( [c7, o3] ), NF*4, ks )
o8 = conv_block( Concatenate()( [c8, o2] ), NF*2, ks )
o9 = conv_block( Concatenate()( [c9, o1] ), NF*1, ks )

out = Activation( AC )( Conv2D( n_slices, 1 )( o9 ) )
att_model = Model(inp, outputs=out )
\end{lstlisting}
\end{lrbox}
\noindent 
\colorbox{gray!10}{
\setlength{\fboxrule}{0ex} 
\fbox{%
  \begin{minipage}{\dimexpr\linewidth-0ex}
  \usebox{\codebox}    
  \end{minipage}%
  }
}

\begin{lrbox}{\codebox}
\begin{lstlisting}[label={lst:resunet},caption={Select network components highlighted in the main text of Section 1.1},language=python,basicstyle=\small]
 
def residual_block( x, NF, KS, strides=(1,1) ):

  x = Conv2D( NF, KS[0], padding='same', strides=strides )(x)
  x = BatchNormalization()(x)
  x = Activation( 'relu' )(x)

  o = stack_bn_act( x, NF, KS[1:], strides=(1,1) )    # one less 
  x = Add()( [o, x] )  
  return x    

NC=2
ks = [3]*NC

o1 = residual_block( inp, NF*1,  ks ); p1= MaxPooling2D( (2,2) )( o1 )
o2 = residual_block(  p1, NF*2,  ks ); p2= MaxPooling2D( (2,2) )( o2 )
o3 = residual_block(  p2, NF*4,  ks ); p3= MaxPooling2D( (2,2) )( o3 )
o4 = residual_block(  p3, NF*8,  ks ); p4= MaxPooling2D( (2,2) )( o4 )
o5 = residual_block(  p4, NF*16, ks );

o6 = concatenate( [deconv_block( o5, NF*8, 2, strides=(2,2) ), o4])
o6 = residual_block( o6, NF*8, ks )
o7 = concatenate( [deconv_block( o6, NF*4, 2, strides=(2,2) ), o3]) 
o7 = residual_block( o7, NF*4, ks )
o8 = concatenate( [deconv_block( o7, NF*2, 2, strides=(2,2) ), o2]) 
o8 = residual_block( o8, NF*2, ks )
o9 = concatenate( [deconv_block( o8, NF*1, 2, strides=(2,2) ), o1]) 
o9 = residual_block( o9, NF*1, ks )

... # same as Code Listing 1        
\end{lstlisting}
\end{lrbox}
\noindent 
\colorbox{gray!10}{
\setlength{\fboxrule}{0ex} 
\fbox{%
  \begin{minipage}{\dimexpr\linewidth-0ex}
  \usebox{\codebox}    
  \end{minipage}%
  }
}

\section{Methods}
\label{sec:methods}


\subsection{Materials}

A training set of 90 ultrasound volumes were provided by the challenge organizers. Each scan captures piece(s) of a steel pipe, potentially containing artifacts inside these pipes. Corresponding Surface mesh of the pipe (pieces) were created by an ``experienced data analyst'' \cite{challenge}. Five of these surface meshes (corresponding to volumes 1 to 5) were provided to the challenge participants and herein referred to as reference masks. Figures 1-5 show examples of the surface renderings of these reference meshes. As the reference labels for the remaining 85 volumes were not provided at the time of challenge, these volumes were mainly used in this study as test samples.

\subsection{Preprocessing}

Each of the reference mesh was first encoded into an image representation. To do so, the vertices of each mesh were read into memory. Next, a binary mask was created to encode the mesh vertices, taking  into account the voxel spacing of the ultrasound volumes. To facilitate learning, the point cloud mask was dilated so that the edges of the binary mask softens. An alternative approach would be to apply Gaussian blur \cite{smrfi} but we found this simpler approach sufficient and computationally more efficient. 
\\\\
To read in the corresponding ultrasound volume, a meta file was created for each of the raw ultrasound files (example in Sec. \ref{sec:meta}; script is also provided under \url{https://github.com/lisatwyw/smrvis/blob/main/utils/write_meta.sh}) and Python package \code{SimpleITK} was used (example code snippet follows). To ensure that the fitted models would be robust to noise, we only preprocessed the input ultrasound data with two steps: down-sampling the image resolution and rescaling their intensity values to [0,1].

\begin{lrbox}{\codebox}
\begin{lstlisting}[label={lst:sitk},caption={Code listing for reading data using SimpleITK and Plydata},language=python,basicstyle=\small]
import SimpleITK as sitk

vols={}
i = 1 # sample_id 
filename='train_data/training/volumes/scan_%03d.mhd'% i

header=sitk.ReadImage(filename)
vols[i]=sitk.GetArrayFromImage(header)

from plyfile import PlyData
plydata,verts,faces={},{},{}

filename='train_data/training/meshes/scan_%03d.ply' % i
vx = plydata[i]['vertex']['x']
vy = plydata[i]['vertex']['y']
vz = plydata[i]['vertex']['z']
verts[i] = [ (vx[d],vy[d],vz[d]) for d in range(len(vx)) ]                
num_faces= plydata[i]['face'].count            
faces[i] = [ plydata[i]['face'][d][0] for d in range(num_faces) ]   
\end{lstlisting}
\end{lrbox}

\noindent 
\colorbox{gray!10}{
\setlength{\fboxrule}{0ex} 
\fbox{%
  \begin{minipage}{\dimexpr\linewidth-0ex}
  \usebox{\codebox}    
  \end{minipage}%
  }
}

\begin{lrbox}{\codebox}
\begin{lstlisting}[language={python},label={lst:pyvista},basicstyle={\small},caption={Code listing for reading in mesh data and generating their screenshots with Pyvista.}]
import pyvista 
# we need to start the frame buffer even if plotting offline
pyvista.start_xvfb()
plotter = pyvista.Plotter(off_screen=True)
mesh = pyvista.read(ply_filename)
plotter.add_mesh( mesh, opacity=.3, color='grey' )
plotter.add_title('Estimated mesh for volume#%d'%i)
plotter.show( screenshot = out + '_screenshot.png')
\end{lstlisting}
\end{lrbox}

\noindent
\colorbox{gray!10}{
\setlength{\fboxrule}{0ex}
    \fbox{%
      \begin{minipage}{\dimexpr\linewidth-0ex}
      \usebox{\codebox}    
      \end{minipage}%
    }
}


\begin{figure}
    \centering
   
    \begin{subfigure}[b]{0.4\textwidth}
        \centering
        \includegraphics[width=\textwidth]{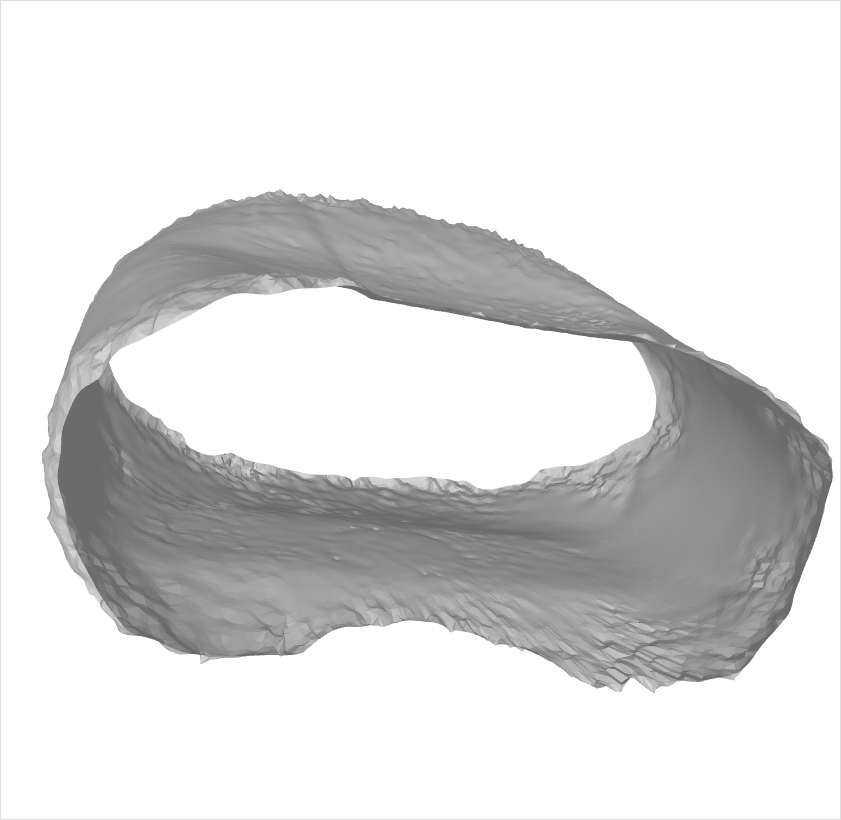}
        \caption[Network2]%
        {{\small Reference mesh of volume 1}}    
        \label{fig:mean and std of net14}
    \end{subfigure}
    \hfill
    \begin{subfigure}[b]{0.4\textwidth}  
        \centering 
        \includegraphics[width=\textwidth]{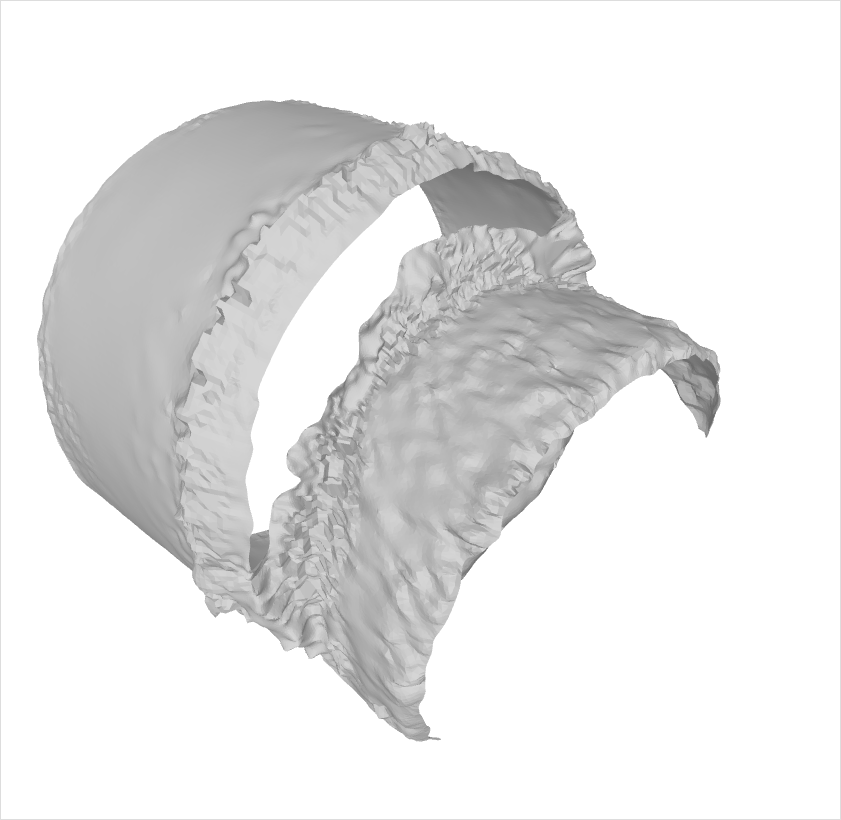}
        \caption[]%
        {{\small  Reference mesh of volume 2}}    
        \label{fig:mean and std of net24}
    \end{subfigure}
    \hfill
    \vskip\baselineskip
    \begin{subfigure}[b]{0.4\textwidth}   
        \centering 
        \includegraphics[width=\textwidth]{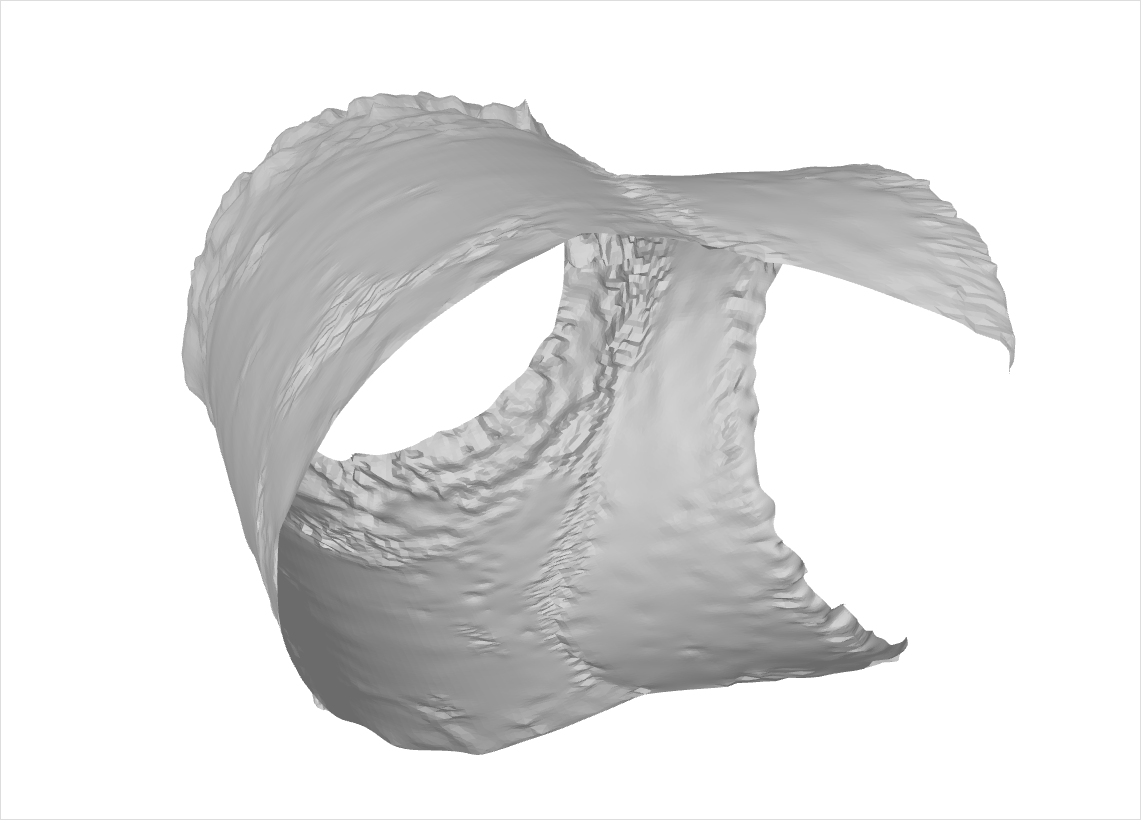}
        \caption[]%
        {{\small  Reference mesh of volume 3}}    
        \label{fig:mean and std of net34}
    \end{subfigure}
    \hfill
    \begin{subfigure}[b]{0.4\textwidth}   
        \centering 
        \includegraphics[width=\textwidth]{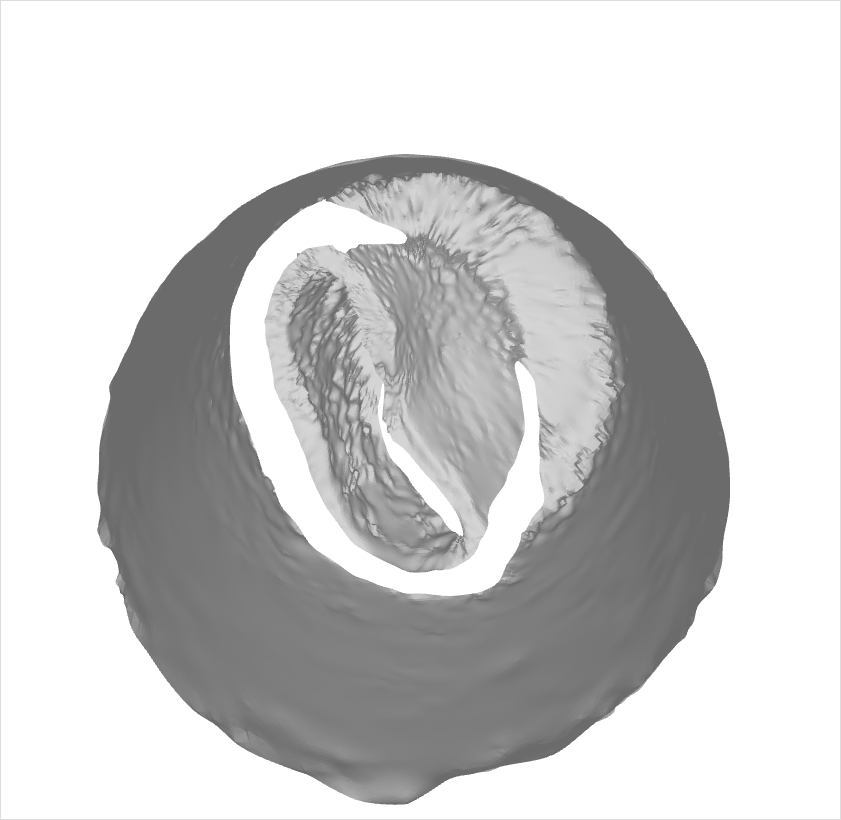}
        \caption[]%
        {{\small  Reference mesh of volume 4}}    
        \label{fig:mean and std of net44}
    \end{subfigure}
    \caption[]
    {\small Reference meshes provided by the challenge organizers.} 
    \label{fig:meshes}
\end{figure}

\begin{figure*}[t]\centering
\includegraphics[width=.35\textwidth]{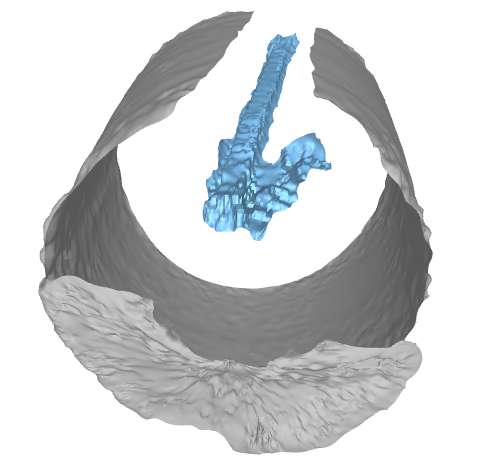}
\caption{Surface rendering of a pipe (grey) containing artifacts (blue).}
\label{fig:m5}
\end{figure*}



\subsection{Models}

Based on observations and evidence from prior studies \cite{Kugelman,tang2020}, we elected to explore the following variants of the U-Net: W-Net (Wnet), the recurrent-residual U-Net (R2-Unet), Squeeze and Excitation U-Net (SE-Unet), U-Net++, and an Attention U-Net (Att-Unet).


\subsection{Training setup}\label{sec:setup}

We divided the provided labeled meshes with three non-overlapping subsets: validation set is used for early-stopping using slices from one volume while the test set is constructed from another volume for evaluation of the training progress, with the remaining volumes all used for training. For instance, a trial might involve volume 5 as the test volume, volume 4 as the validation set, and volumes 1-3 as the training set.

\subsubsection{Data augmentation}

We augmented the training dataset by perturbing the input training set on-the-fly with random translations, rotations, and mirroring on the x- and y-axes. We examined two approaches: applying these different types of transforms simultaneously, or exclusively. 
We also explored the effects of a training schedule wherein training volumes that have been reoriented entirely are only presented at a later phase. 


\subsubsection{Optimization}

We employed the Adaptive Moment Estimator (ADAM) optimization algorithm with default parameters and explored values of 0.0001, 0.008, 0.001, and 0.1 as the initial learning rate in the context of three learning rate scheduling schemes, namely, cyclical decay, cosine decay, and polynomial decay \cite{poly}.

Training was permitted to run for 300 epochs or terminated early when there is no reduction in the metrics computed on the validation set. We empirically explored the use of Dice and Jaccard as the validation metric but found training diverged when these  two metrics were used, most likely due to scarcity of mesh vertices in relation to the its enclosing volume.  We thus elected to use the same loss function (but computed on the validation set) for determining the termination criteria. Example progress plots are shown in the Appendix.




\subsection{Ablation studies conducted}


We conducted ablation trials to answer design questions surrounding the following components and present comparisons in the Results section and the Appendix. 

\begin{enumerate}
    \item Image resolution (IR): how might the image resolution used to train the models affect accuracy? (According to a 2019 review \cite{viz3d},  ultrasound scanners typically acquire data that will be reconstructed to matrix size of 512 $\times$ 512 with 256 intensity levels. Hence, we explored the standard resolution of 256 $\times$ 256 and the non-standard resolution of 384 $\times$ 384, which correspond to down-sampling factors of three and two, respectively); 
    \item Activation function (AC): prior work advocate \cite{bce} the use of sigmoid, hard sigmoid, and hyperbolic tangent function based on datasets involving magnetic resonance and computed tomography data; do results generalize to ultrasonic data?
    \item Loss function (LS): previous studies \cite{bce,loss,activations23} have observed inter-plays between the loss function and activations for image classification; we hypothesize that their results may not generalize to ultrasound point extraction and hence explore Dice, binary cross entropy (BCE) and binary focal cross entropy loss terms (BFCE); 
    \item Selection scheme (SE): should all ultrasound sections be presented to the model during training (SE=1) or only sections containing the reference mesh (SE=2) be presented to ease training? 
    \item Random transform (RT): how the type(s) of random transformations affect the training progress? e.g. should all random transforms be allowed or only one type at a time?
    \item Mesh encoding (EN): how should the mesh vertices be encoded in the image space? Would attenuating boundaries of the surface mesh help training when this encoding strategy is used in conjunction with alternative loss functions such as pixel-wise mean squared difference or absolute difference?  
    \item Number of filters (NF): would reducing the number of filters from the default size of 16 to 8 impact performance severely?    
\end{enumerate}

\subsection{Evaluation metric and model selection}

Following the Challenge's evaluation protocol, we employ Chamfer distance (CD) measure, which is defined as:
\begin{equation}\label{eq:cdist}
    CD( \mathcal{S}, \mathcal{ T} )= \frac{1}{ |\mathcal S|} \sum_{x \in S} min_{y \in T} || x - y ||^2_2 + \frac{1}{ |\mathcal T |} \sum_{y \in \mathcal T} min_{x \in S} || x - y ||^2_2 
\end{equation} where $\mathcal S$ and $\mathcal T$ denote source and target point clouds, respectively. 
\\\\
To enable computation without needing a graphics card, we approximated the distance using $v=$10,000 points randomly drawn from each point cloud as we find the measured Chamfer distance to be relatively stable for this choice of $v$.

\subsection{Mesh surface extraction}
We employ Python packages \code{Veko} and \code{Pyvista} to respectively extract isosurfaces and  visualize the extracted isosurfaces for volume rendering of the mesh surfaces (code listing \ref{lst:pyvista}). Figure 1-5 illustrate reference meshes rendered for volumes 1-5.

\subsection{Implementation and deployment details}
 
As mentioned earlier, opensource code for the U-Net variants \cite{Kugelman} were adapted so that different activation functions could be tested.
\\\\
All experiments were conducted in a virtual environment with Python 3.10, Tensorflow 2.12 and Torch 1.13.0. Graphical processing unit (GPU) cards explored include  NVIDIA Tesla V100, Tesla T4, and P1000-SXM2 (CUDA Version 12.0).
\\\\ 
To maximize reproducibility \cite{docker}, source scripts will be updated on the repository with human- (and machine-) readable instructions
for emulating the computing environment needed to run these scripts in the form of a docker container at \url{https://github.com/lisatwyw/smrvis/}. 

\section{Results}

\begin{figure*}[t]\centering
\includegraphics[width=.9\textwidth]{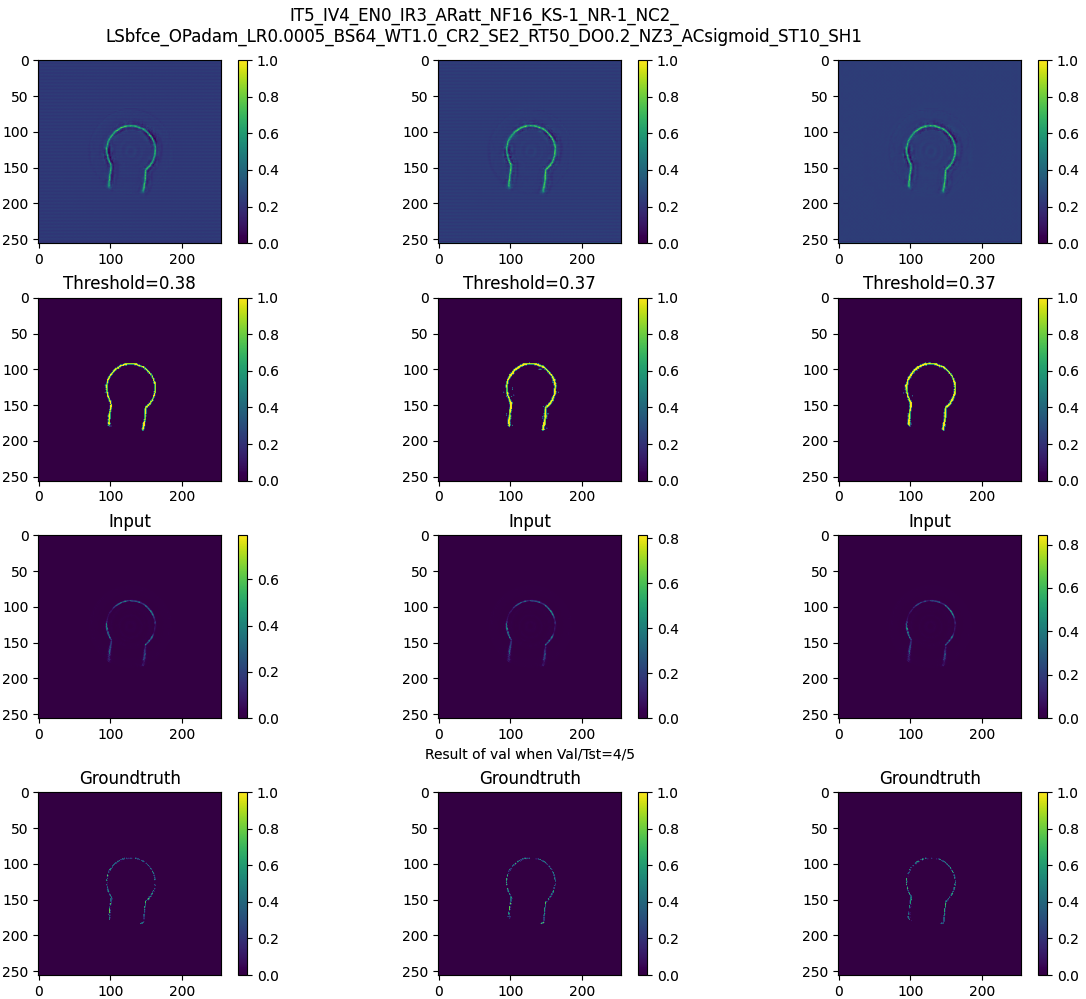}
\caption{Example 3-slice input to each model. Top to bottom: probabilistic output, thresholded output, input ultrasound slice, and its corresponding reference mesh label.}
\label{fig:test442d}
\end{figure*}

When each ultrasound volume of 1281 slices was down-sampled to an axial resolution of 256 $\times$ 256, inference with a single GPU with V100 and single-core CPU for instance took less than one minute and 431.2 $\pm$0.9 seconds (about 7 minutes), respectively.
\\\\
Figure \ref{fig:pc1}-Figure \ref{fig:pc5} each illustrates the point cloud extracted from the training volumes by the proposed framework and the point clouds provided in the reference mesh files.
Figure \ref{fig:test442d} visualizes results in two dimensions, while Figure \ref{fig:test48} presents the isosurface computed from a randomly selected test volume.
\\\\
We next present quantitative results with Table \ref{tab:res1}-\ref{tab:res4} that generally report the Chamfer distances between the reconstructed and reference point clouds.
\\\\
More specifically, Table \ref{tab:res1} presents trials that failed to converge while tables \ref{tab:res2}-\ref{tab:res4} present trials that yielded satisfactory Chamfer distances (below 95.0). In generating these tables, samples from select volumes were used as the training and validation set while a left out volume was used as a test (unseen) sample as described in Section \ref{sec:setup}; these are marked in the tables as `(val)' and `(tst)' to denote validation and test set, respectively.
\\\\
Based on the results from the trials summarized by these tables, volumes 4 and 5 had the lowest and highest Chamfer distance, respectively. This may be explained by  Figure \ref{fig:m5}, which shows that volume 5 captured a pipe with an object inside, rendering an obstacle to achieving low Chamfer distances between the extracted and reference point clouds.
\\\\
Results from Table \ref{tab:res3} suggest that the choice on the encoding schemes of the reference mesh labels did not impact performance in significant ways. Empirical results (Appendix) suggest that taking the average of contiguous slices maybe superior to taking the solution of the most confident slice.
\\\\
Table \ref{tab:params} lists the hyper-parameters of the configurations tested for the U-Net variants and example statistics on training convergence.
\\\\
More results are presented in the Appendix; in short, we did not find significant difference between the following:
\begin{enumerate}
\item $384 \times 384$ (IR=2) vs $256 \times 256$ (IR=3);
\item Sigmoid, hard sigmoid, and hyperbolic tangent will lead to slightly different effective sizes \cite{benchmark22}, which will impact the choice of threshold values;
\item We did not find improvement of accuracy due to the inclusion of Dice score nor did we find BFCE to be superior over BCE; 
\item Use of 8  filters and 16 filters gave comparable results 
\end{enumerate}
Conversely, the following affected the success of training:
\begin{enumerate}
\item Chance of training success was increased when the encoded reference meshes were dilated; 
\item Chance of training success increased when the second scheme was adopted for sample selection (omit training samples that did not contain the mesh and thus less relevant); 
\item Simultaneous application of different random transformations may render training more difficult; we found that a progressive scheme of introducing multiple transformations only in a later phase of training to be an effective solution. 
\end{enumerate}
In summary, the models did not appear to have over-fitted to  the training set as including the hardest sample (volume 5) did not lead to reduction in error. There appears to be no significant difference due to the choice of model, 
image resolution, and random transformation schemes (more results presented in the Appendix). We observed that training diverged when reoriented volumes were presented too early during model training.

\begin{table}[h]
\begin{center}\small
\begin{tabular}{l |   l | l r | l l | l l  } 
 \hline 
Model (AR) & \# of filters (NF) & \# of layers & Model size & Training set  & activation (AC) & \# of epochs & $t$ \\ \hline 
Att-Unet & 16 & 190 &  1,989,767 &2,3,4 [1/5] &  sigmoid & 54 & 0.31 \\
 & 8 & & 1,989,767  & All & sigmoid & 51 & 0.32 \\  
R2-Unet & 16 &  158 &  1,999,571  &3,4,5  [2/1] & sigmoid &35 & 0.32 \\
& 8, NR=NC=3 &356& 751,035 &  All & sigmoid & 49 &  0.42 \\
U-Net++ & 8  & 242 & 514,219  &2,4,5 [3/2] & sigmoid & 167 & 0.41 \\  
& 16 &  242& 2,050,387 & 1,2,3 [5/4] & sigmoid & 121  & 0.39 \\ 
SE-Unet & 16  & 161 &  2,054,355 & 1,2,5 [4/3] & tanh &  30 & 0.40 \\
& 8 & 161 & 515,179  & All & tanh & 36 & 0.40 \\
W-Net & 16 & 168 &  1,159,091 &1,2,3 [5/4] &  hard sigmoid  & 100 & 0.34\\ \hline 
\end{tabular}
\caption{Number of trainable parameters of the explored model configurations is shown under column `Model size'. The last column records the threshold $t$ applied to the probabilistic output in order to generate isosurfaces of the predicted mesh surface. \label{tab:params} } 
\end{center}
\end{table}

\begin{table}[ht]
\begin{center}\small
\begin{tabular}{l | l | l l  l   l   l } 
 \hline
Model & Settings & Volume 1 & Volume 2 & Volume 3 & Volume 4 & Volume 5  \\ \hline
- & - & 167.6  & 167.6  & 167.6  & 167.6  & 167.6  \\
R2Unet & BCE  & 105.8  & 103.8  & 102.7  & 105.8  (val) & 118.6 (tst)   \\
SE-Unet & BFCE & 107.9  &    108.1  &    102.3  &    104.8 (val)  &    114.9  (tst)  \\ 
R2Unet & Hard sigmoid/ 2-1  ($t$=0.32) &    147.4 (tst) &    137.6 (val)  &    144.4  &    134.1  &    149.8   \\
\hline
\end{tabular}
\caption{Baseline for subsequent evaluations based on Chamfer distance. For reference, treating every voxel position as part of the reconstructed mesh yields the worse possible distance as 167.6. \label{tab:res1} } 
\end{center}
\end{table}

\begin{table}[hb]
\begin{center}\small
\begin{tabular}{l | l | l |  l l  l   l   l } 
 \hline
Model & Settings & $t$ & Volume 1 & Volume 2 & Volume 3 & Volume 4 & Volume 5  \\ \hline

Att-Unet& Trained on 1,2,3 & 0.19 &  84.0  & 75.1  & 75.9  & 68.4 (val) & 94.0 (tst) \\
 && 0.28 & 82.1  & 73.4  & 72.5  & 66.3  (val)& 93.2 (tst)\\ 
 && 0.37 & 83.2  & 73.7  & 70.5  & \bcell 66.1  (val) & \bcell 93.2 (tst)\\ \hline

R2-Unet & Trained on 3,4,5 & 0.15 &    81.5 (tst) &    73.3  (val) &    73.3  &    65.1  &    91.8  \\  
& & 0.22 &    \bcell 77.9 (tst) &   \bcell  73.1  (val)&    72.3  &    65.3  &    91.8  \\  \hline
& Trained on 1,4,5 & 0.2 &    83.1  &    74.6 (tst)  &    75.0 (val)  &    66.4  &    93.4  \\  
& & 0.3 &    80.4  &    73.7 (tst)   &   73.5 (val) &    65.8  &    93.3  \\  
& & 0.4 &    73.8  &  \bcell   73.9 (tst)  &  \bcell  72.1 (val)   &    65.9  &    93.4 \\ \hline
 & Trained on 1,2,3 & 0.14 & 82.0  & 74.8  & 73.0  & 65.5  (val) & 92.1 (tst) \\
 &      & 0.21 & 81.8  & 71.4  & 72.7  & 65.4 (val) & 92.2 (tst)\\
       & & 0.25 & 82.1  & 71.2  & 72.1  & 65.6 (val) & 92.2 (tst)\\
       & & 0.35 & NIL   & 71.7  & 71.3  & \bcell 65.3 (val) & \bcell 91.1 (tst)\\\hline

SE-Unet & Trained on 3,4,5 & 0.2 &    82.0 (val)  &    72.7  &    73.0  (tst) &    65.9  &    92.1  \\
& & 0.3 &    80.3 (val)  &    72.4  &    72.7  (tst) &    65.3  &    92.3   \\ 
& & 0.4 &  \bcell  78.3 (val) &    73.3  &    \bcell 71.1 (tst) &    65.6  &    92.8   \\ \hline
& Trained on 1,2,3 & 0.20 & 81.8  &    72.7  &    71.9  &  \bcell   64.7 (val)  &  \bcell  92.4   (tst) \\
  & & 0.30 & 80.1  &    72.2  &    71.2  &   65.0  (val) &    93.0  (tst)\\   
& & 0.32 & 79.6  &    72.3  &    70.9  &    65.1 (val) &    93.2 (tst)  \\ 
\hline 

Wnet & Trained on 1,2,5 & 0.338 &  72.8  &    72.6  &    72.3  (val) &    65.7 (tst)  &    92.7   \\
& & 0.375 &    71.7  &    72.8  &    72.1 (val)   &    65.9 (tst) &    93.0  \\
& & 0.413 &    69.9  &    73.1  &   \bcell 71.9 (val)  &   \bcell 66.1  (tst)&    93.4  \\ \hline 
\end{tabular}
\caption{Effects of training set on validation and test volumes. Performance evaluation of different networks based on Chamfer distance. For each model, the best test score selected by the lowest validation error (i.e. lowest Chamfer distance) is  highlighted with blue cells.\label{tab:res2}} 
\end{center}
\end{table}

\begin{table}[h]
\begin{center}\small
\begin{tabular}{l| l | l |  l l  l  l l } 
 \hline
Model & Settings & threshold & Volume 1 & Volume 2 & Volume 3 & Volume 4 & Volume 5  \\ \hline
Att-Unet & Saturated mask & 0.2 &  82.1  &    73.0  &    72.7 (val)  &    66.6  (tst) &  92.5 \\
&    & 0.3 &   81.3  &    72.4  &    72.3  (val) &    66.1  (tst) &    92.4   \\ 
&    & 0.4 &  78.9  &    72.5  &   \bcell{70.8  (val)} &  \bcell{65.7  (tst)}&   92.7   \\  \hline 
& Solid mask & 0.22 &  81.1  &   72.7 &    72.4 (val) &    65.7 (tst) &  92.3  \\
&  & 0.33 & 79.2  &  72.8  &    71.6 (val) &    65.5  (tst) &    92.6   \\
&  & 0.37 & 78.8  &  73.1  &  \bcell  71.3 (val) & \bcell 65.5  (tst) &    92.6  \\  \hline 
\end{tabular}

\caption{Comparison of the encoding schemes of the reference mesh labels. \label{tab:res3} } 
\end{center}
\end{table}

\begin{table}[h]
\begin{center}\small
\begin{tabular}{l | l | l | l l  l   l  l } 
 \hline
Model & Settings & $t$ & Volume 1 & Volume 2 & Volume 3 & Volume 4 & Volume 5  \\ \hline
Att-Unet & Hard sigmoid & 0.20 & 116.4  &    107.0  &    106.2  &    95.7  &    114.0  \\
& & 0.30 &  81.4  &    73.8  &    74.3   (val) &    66.4   (tst) &    93.0  \\
& & 0.40  &   78.5  &    73.7  &   \bcell 73.2   (val) &  \bcell  65.9   (tst) &    93.1   \\ \hline \hline 

R2-Unet & Linear & 0.28
&    76.2  &    73.0  &    72.6  &    65.1  &    93.8   \\  \hline \hline 

SE-Unet & Hard sigmoid & 0.18 & 81.9  &    73.3  &    74.6  (val) &    66.8  (tst) &    92.4  \\
 & & 0.27 &  80.3  &    73.0  &    72.9  &    65.8  &    92.7  \\
 & & 0.37 &    81.2  &    74.3  &  \bcell  71.4   (val) &  \bcell  66.1  (tst)  &    93.4   \\ \hline

& tanh & 0.26  
&    79.7  &    73.8  &    71.5  &    66.0  &    93.5  \\
& & 0.29  
&    80.4  &    74.0  &   \bcell 70.9 (val) &   \bcell 66.0  (tst) &    94.4  \\ \hline \hline

Wnet & Hard sigmoid (BFCE) & 0.34  
&    81.6  &    72.7  &    69.6  &    64.3  &    92.4  \\
& 
& 0.37 &  80.1  &    72.8  &   \bcell  69.1 (val)  & \bcell   64.1 (tst) &    92.1  \\

&  Hard sigmoid with BCE & 0.34  
&  81.6  &    72.7  &    69.6  &    64.3  &    92.4   \\
& & 0.37  
&  80.1  &    72.8  &   \bcell 69.1  (val) & \bcell   64.1 (tst)  &    92.1  \\
\hline
\end{tabular}
\caption{Comparisons on the loss and activation functions. \label{tab:res6}} 
\end{center}
\end{table}

\begin{table}[h]
\begin{center}\small
\begin{tabular}{l | l | l | l l  l   l  l } 
 \hline
Model & Settings & $t$ & Volume 1 & Volume 2 & Volume 3 & Volume 4 & Volume 5  \\ \hline
Att-Unet & Max & 0.20
&    81.0  &    74.4  &    73.7  &    66.6  &    93.0  \\
& & 0.30
&    79.9  &    74.0  &    73.3  &   66.3  &    93.0   \\
& & 0.40
&    79.4  &    74.0  & \bcell   73.1 (val)  &  \bcell 66.3 (tst)  &    93.2   \\\hline 
& Mean & 0.40 &  78.9  &    72.5  &   \bcell{70.8  (val)} &  \bcell{65.7  (tst)} &   92.7   \\  \hline  \hline 
SE-Unet & Max &0.33 
&    79.9  &    73.4  &    72.4 (val) &    65.8 (tst) &    93.1  \\
&&0.36 &    80.4  &    73.9  &  \bcell  71.7 (val) & \bcell  65.9 (tst) &    93.4   \\ \hline
& Single slice only & 0.33 
&    80.6  &    73.9  &    72.0 (val) &    65.8 (tst)  &    93.0  \\
& & 0.36  
&    81.2  &    74.4  &  \bcell  71.3 (val) &  \bcell  66.3 (tst)  &    94.2  \\ \hline
 & Mean & 0.37 &    81.2  &    74.3  &  \bcell  71.4   (val) &  \bcell  66.1  (tst)  &    93.4    \\ \hline 

 SE-Unet with tanh & Mean &  0.23
&    77.4  &    73.8  &    71.8  &    65.9  &    93.6  \\
&  & 0.26
&    77.4  &    74.0  &   \bcell 71.2 (val)  & \bcell   66.0  (tst) &    94.8  \\\hline
 & Max & 0.23
&    78.2  &    73.7  &    72.2  &    66.0  &    93.3   \\
& & 0.26
&    78.0  &    73.9  &    71.8  &    66.1  &    93.5   \\\hline 
& Single slice only  & 0.23
&    79.5  &    73.7  &    72.0  &    65.8  &    93.5  \\
& & 0.26
&    79.6  &    73.9  & \bcell   71.5  (val) &  \bcell  66.0  (tst) &    93.5  \\ 
 \hline \hline

W-Net & Max & 0.338 &   89.0  &    80.7  &    84.2 (val)  &    74.5 (tst)  &    98.1   \\
&  & 0.375
&    81.6  &    76.5  &    78.1  (val) &    69.5 (tst) &    94.8  \\
 & & 0.413
&    77.5  &    75.2  &   \bcell  75.4  (val) &   \bcell 67.7 (tst) &    94.6   \\
\hline
& Single slice only & 0.338  
&    82.9  &    77.3  &    79.4  &    70.2  &    95.8   \\
& & 0.413  
&    77.5  &    76.5  &  \bcell  77.5 (val)  &  \bcell  68.6 (tst) &    95.2  \\  \hline 
& Mean (from Table \ref{tab:res2}) & 0.413 &    69.9  &    73.1  &   \bcell 71.9 (val)  &   \bcell 66.1  (tst)&    93.4  \\ \hline

\end{tabular}

\caption{Comparisons on the choice of aggregation scheme. Selection based on the validation set suggests that aggregation by mean achieves distance of 66.1 on the test volume, which is a slight improvement compared to alternative choices (67.7- 68.6) for the Wnet. For the SE-Unet, there is no distinctive impact from the aggregation choice (Chamfer distance ranged from 65.8-66.3). \label{tab:res4} } 
\end{center}
\end{table}

\section{Discussions}\label{sec:discuss}

While the 2022 benchmark study \cite{Kugelman} hinted a trade-off between memory requirements, training time, and accuracy between the U-Net variants, our study did not observe an obvious difference between the U-Net variants in terms of accuracy \emph{nor} training time, potentially due to the small number of labeled volumes. 
Nevertheless, the use of eight sets of filter in U-Net++ did not increase error, thereby suggesting room for optimizing the hyper-parameters that would best balance between model widths and depths without compromising accuracy. 
\\\\
Recall that the two key factors that render CNNs computationally expensive include the number of model parameters and the number of multiply-add (MAD) operations required by each. Depth-separable and group convolutions are two strategies to balance between the number of model parameters and the number of MAD operations needed. To optimize both factors, Liu et al. proposed ConvNeXt that employs convolutional blocks composed of depth-separable convolution followed  by a network component called \textbf{inverted bottleneck}  and \textbf{point-wise convolutions}. Following this line of design thinking, Heinrich and Hagenah \cite{heinrich23} very recently (2023) propose two orthogonal strategies specifically to lower the computational requirements of the self-configured, supervised learning framework known as the ``no new U-Net`` (nnU-Net) \cite{nnunet}. Firstly, \textbf{partial convolution} uses ``T-shaped'' spatial convolution previously proposed by Chen et al. \cite{chen23}  to perform spatial convolution only on select channels and subsequently applies point-wise operators on the remaining channels within the {inverted bottleneck} component. Secondly, \textbf{re-parameterization} allows one to reduce the size of the model after it has been trained. This was made possible by placing \textbf{batch normalization} between the first and second convolution blocks and dropping the use of non-linear activation functions \cite{heinrich23}. According to  the presented experimental results \cite{heinrich23}, these two strategies led to reduction of model sizes by a factor of 3 to 4 and shorter inference time of about twice as fast when compared to the original version of nnU-Net. 
\\\\
The objectives of the present study did not involving placing restrictions on memory consumption, computational demands, and inference times. Future researchers may further tackle these constraints by exploring and evaluating these latest state-of-art extensions of U-Net \cite{nnunet,heinrich23}.
\\\\
Due to time constraints, a major limitation of the present framework is failure to leverage the 85 unlabeled ultrasound volumes provided by the workshop challenge. Future work will quantify the potential advantages of including pseudo labels into our framework and/or the use of other reinforcement learning techniques. Other strategies that generate point clouds directly from data \cite{pointe} could also be examined.

\section{Conclusion}

In this brief note, we explored the feasibility of surface mesh reconstruction via point cloud estimation as an image to mask generation problem. This initial framework opens door to possibility of leveraging (pretrained) deep and wide networks published in the wild. The source code developed during the course of this experimental prototyping period will be posted at \url{https://github.com/lisatwyw/smrvis}. We hope the research communities will find this quick prototype consisting of a few Python scripts approachable.
\\\\
\noindent\textbf{\large{Acknowledgements}}
\\\\
We sincerely thank DarkVision Technologies Inc. for provision of the ultrasound dataset and hosting this exciting challenge. The author also expresses deep gratitude to Tong Tsui Shan and Kim Chuen Tang for their support.

\newpage 
\appendix

\begin{table}[h]
\begin{center}\small
\begin{tabular}{l  l l   l  l l l } 

Model & &&&&& ($f$ to compute $t$, AG)\\ \hline
Att-Unet
&    77.5 (tst)  &    72.9  &    73.4  &    65.4  &    92.8  (val)  &(0.9, 0)\\
&   \bcell 77.1  &    71.9  &    73.4  &    64.5  &   \bcell 92.1&  (1, 0)\\
&    79.3  &    73.7  &    73.3  &    66.2  &    93.0 & (0.9, 1)\\
&    78.9  &    73.3  &    73.3  &    65.9  &    92.7  &(1, 1)\\
&    76.9  &    73.2  &    73.4  &    65.9  &    92.9  &(0.9, 2)\\
&    75.7  &    72.7  &    73.4  &    65.7  &    93.0 & (1, 2) \\

\\
\\
R2-Unet
&    81.7   &    70.9  &    72.4  &    65.6  (val/tst) &    92.3 & (0.9, 0) \\
&    82.8  &    71.1  &    72.1  &    65.6  &    92.1 & (1, 0)\\
&    81.7  &    71.7  &    72.5  &    65.4  &    92.5 & (0.9, 1)\\
&    81.8  &    72.1  &    72.2  &    65.6  &    92.6 & (1, 1)\\
&    82.3  &    70.5  &    72.9  &    65.4  &    92.1 & (0.9, 2)\\
&   \gcell 83.1  &  \gcell  70.8  & \gcell   72.7  &  \gcell  65.4  &  \gcell  91.9 & (1, 2)\\
\\
SE-Unet
&    77.4  &    73.8  &    71.8 (val) &    65.9  (tst)&    93.6  &(0.9, 0)\\
&    77.4  &    74.0  &   \bcell 71.2  (val)&  \bcell  66.0 (tst)  &    94.8  &(1, 0)\\
&    78.2  &    73.7  &    72.2  &    66.0  &    93.3  &(0.9, 1)\\
&    78.0  &    73.9  &    71.8  &    66.1  &    93.5  &(1, 1)\\
&    79.5  &    73.7  &    72.0  &    65.8  &    93.5  &(0.9, 2)\\
&    79.6  &    73.9  &    71.5  &    66.0  &    93.5  &(1, 2)\\

& \gcell 74.4 & \gcell 73.8 &\gcell 71.2 & \gcell 66.4 & \gcell 94.2 \\
\\
U-Net++; NF8/SE2/IR3 
&    77.4  &    72.8  &    72.6 (tst) &    65.1 (val) &    92.4 & (0.9, 0)\\ 
&    75.4  &    73.0  &    72.4  &    65.5  &    92.6  &(1, 0) \\ 
&    78.5  &    72.9  &    72.6  &    65.3  &    92.5  &(0.9, 1)\\ 
&    76.3  &    73.1  &    72.4  &    65.4  &    92.6  &(1, 1) \\ 
&    78.3  &    72.8  &  \bcell  72.6 (tst) &    \bcell 64.9 (val) &    92.4  &(0.9, 2) \\
&    76.1  &    72.9  &    72.5  &    65.4  &    92.5  &(1, 2) \\
\\ 
U-Net++; NF16/SE1/IR2/RT=60  
&    82.6  &    74.0  &    70.9  &    66.3  &    93.4  &(0.9, 0)\\
&    85.4  &    74.4  &    67.9  &    66.9  &    93.4  &(1, 0)\\
&    82.4  &    74.0  &    71.5  &    66.1  &    93.5  &(0.9, 1)\\
&    83.8  &    74.4  &    69.6  &    66.7  &    93.4  &(1, 1)\\
&    82.1  &    74.0  &    71.7  &    \bcell 66.0 (val)  &  \bcell  93.3 (tst) &(0.9, 2)\\
&    83.1  &    73.9  &    70.6  &    66.5  &    93.5  &(1, 2) \\
\\
&    82.4  &    73.4  &    72.7  &    65.9  &    93.2  & (0.7, 0)\\
&    82.3  &    73.7  &    72.1  &    65.9  &    93.3  & (0.8, 0)\\
&    82.3  &    73.4  &    72.9  &    65.8  &    93.2  & (0.7, 1)\\
&    82.3  &    73.7  &    72.3  &    65.9  &    93.5  & (0.8, 1)\\
&    82.7  &    73.4  &    72.8  &    65.9  &    93.2  & (0.7, 2)\\
&    82.4  &    73.7  &    72.3  &    65.8  &    93.3  & (0.8, 2)\\

\\ 
Att-Unet; SE=1; IR=2; RT=45 
&    80.5  &    73.9  &    72.0  &  \bcell  65.9  (val) & \bcell   93.7 (tst) &(0.9, 0)\\
&    81.4  &    74.4  &    70.6  &    66.5  &    94.9  &(1, 0)\\
&    80.5  &    73.9  &    72.5  &    66.0  &    93.4  &(0.9, 1)\\
&    80.6  &    74.3  &    71.4  &    66.3  &    94.7  &(1, 1)\\
&    81.9  &    74.1  &    71.8  &    66.1  &    93.5  &(0.9, 2)\\
&    82.0  &    74.5  &    70.9  &    66.4  &    93.8  &(1, 2) \\

\\
Wnet
&    82.3  &    72.4  &    70.2  &    64.4  (val) &    92.8 (tst) &(0.9, 0)\\
&    81.9  &    72.4  &    69.7  &    \bcell 63.9 (val)  &  \bcell  92.7 (tst)  &(1, 0)\\
&    82.6  &    72.6  &    70.3  &    64.7  &    92.8  &(0.9, 1)\\
&    82.4  &    72.6  &    69.9  &    64.5  &    92.7  &(1, 1)\\\hline 
\end{tabular}
\caption{Evaluation by Chamfer distance computed between the reconstructed and reference point clouds extracted from each ultrasound volume. \label{tab:res7} } 
\end{center}
\end{table}

\section{More example visualizations}\label{sec:viz}

\begin{figure*}[bh]\centering
\includegraphics[width=.5\textwidth]{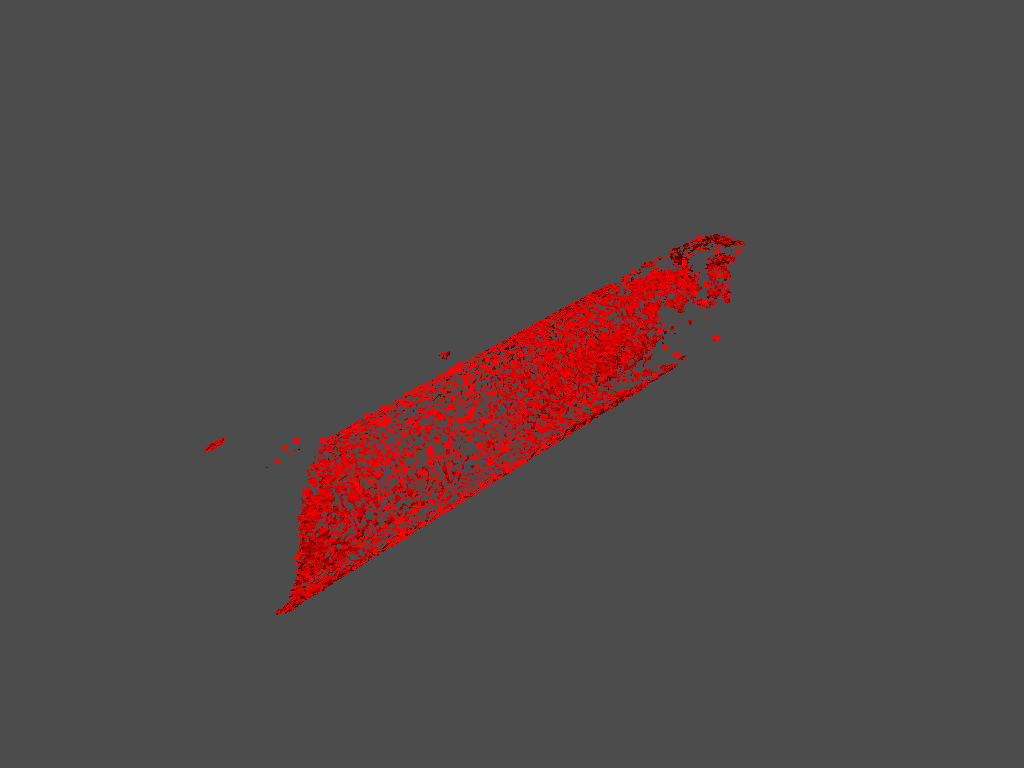}
\caption{Example of a degenerate case.}
\label{fig:degen}
\end{figure*}

\begin{figure*}[t]\centering
\includegraphics[width=.9\textwidth]{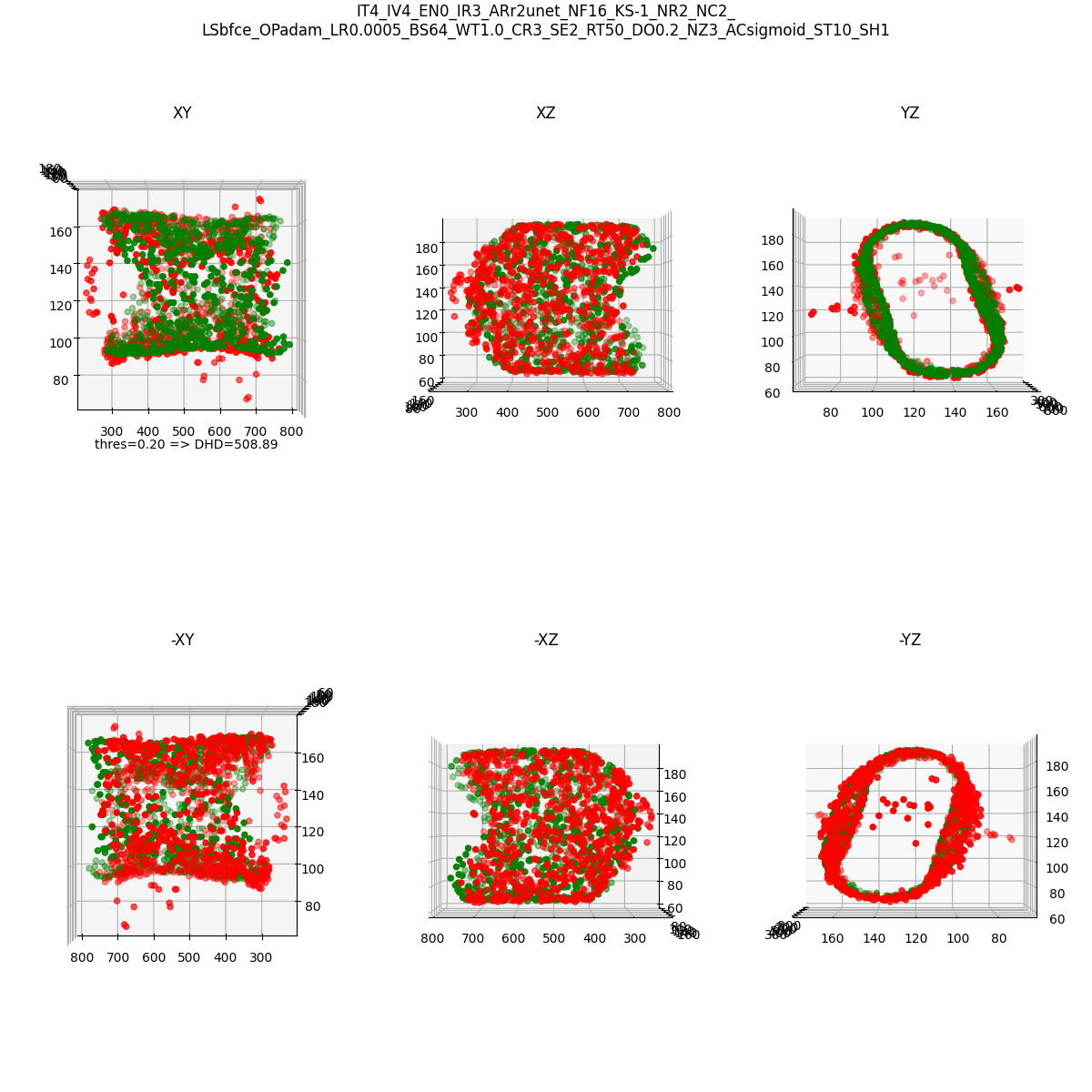}
\caption{Results for volume 1. Point cloud of the reference mesh label (in green) and the point cloud extracted by the proposed framework using a R2-Unet (in red).}
\label{fig:pc1}
\end{figure*}

\begin{figure*}[t]\centering
\includegraphics[width=.9\textwidth]{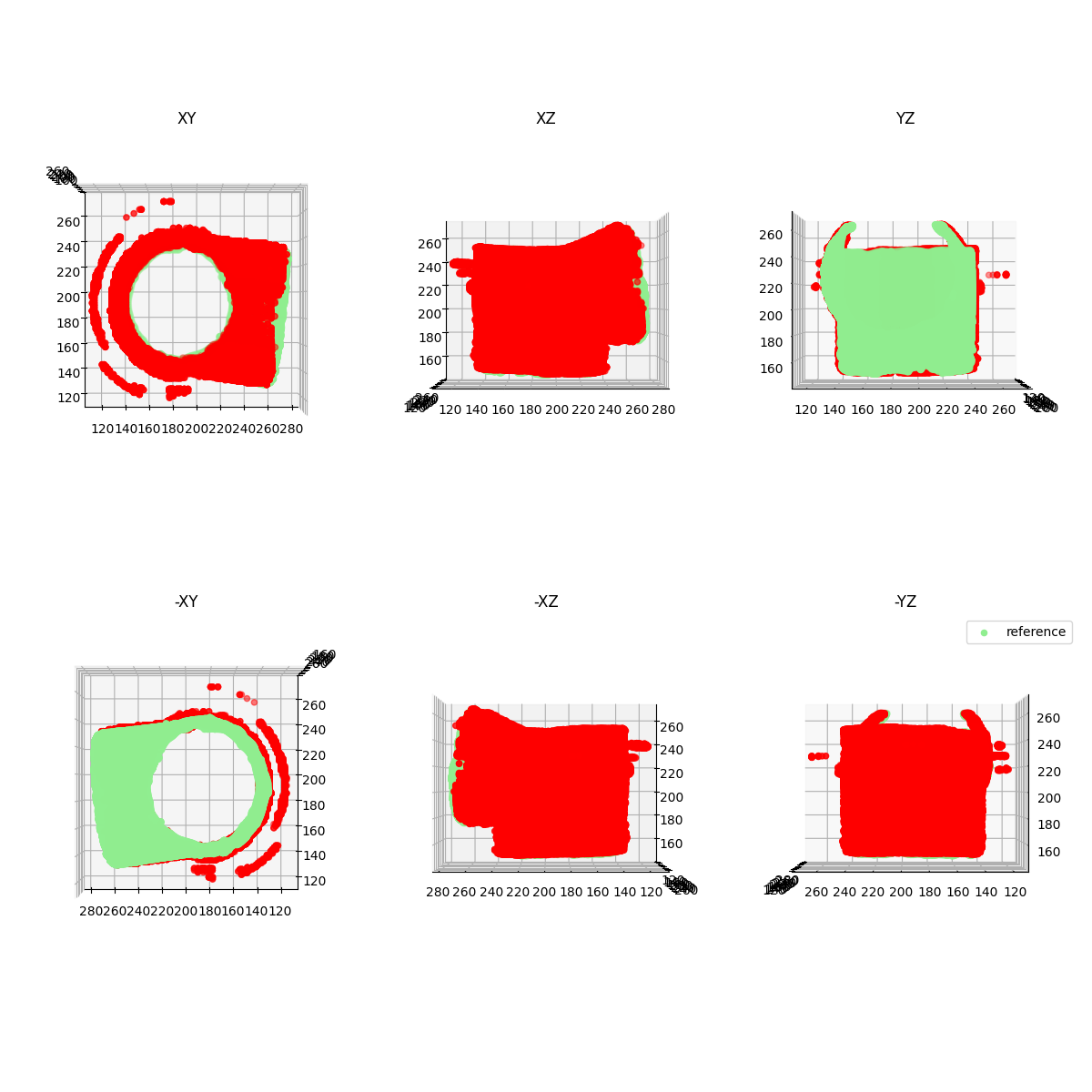}
\caption{Results for volume 3.}
\label{fig:pc3}
\end{figure*}

\begin{figure*}[t]\centering
\includegraphics[width=.9\textwidth]{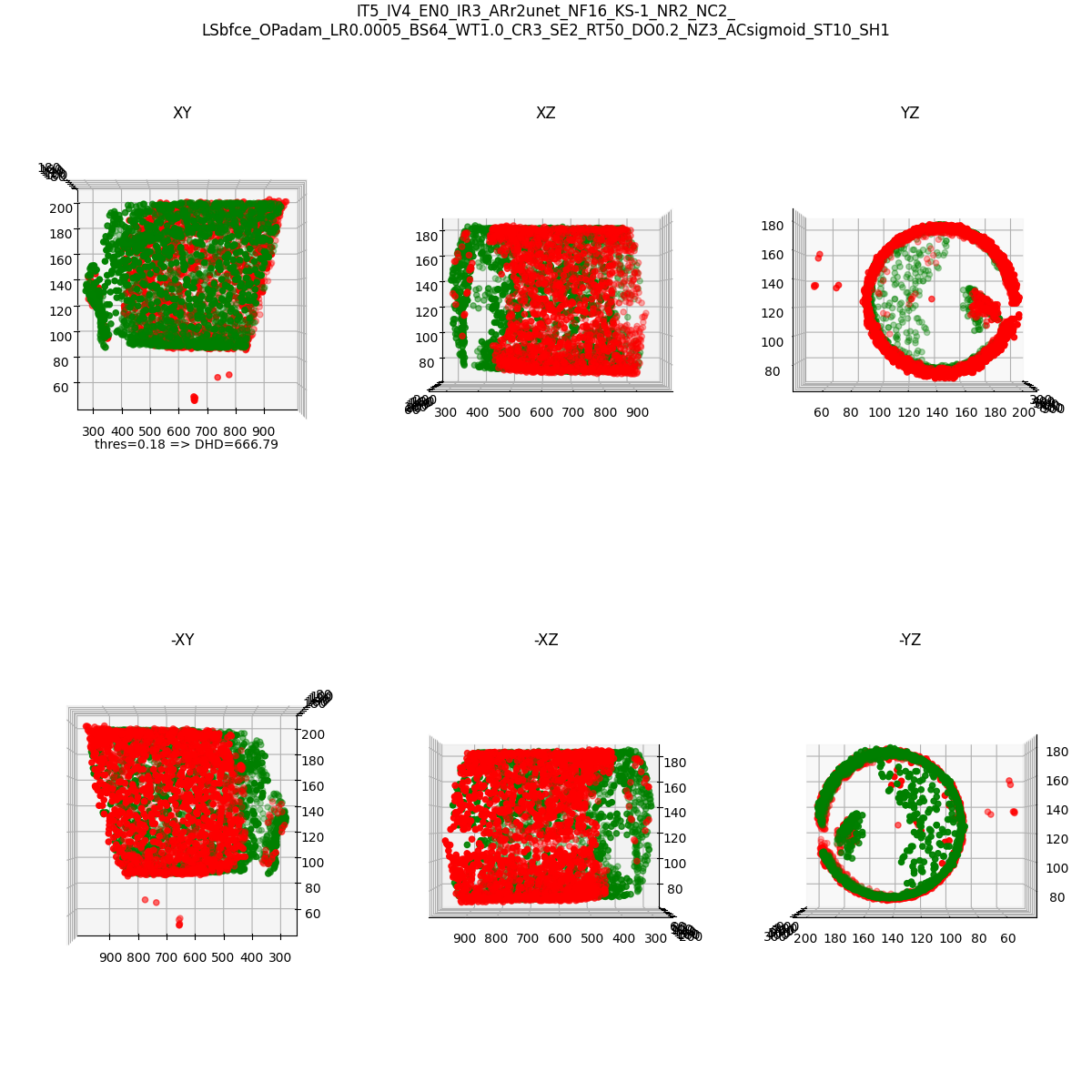}
\caption{Results for volume 4 using an SE-Unet.}
\label{fig:pc4}
\end{figure*}

\begin{figure*}[t]\centering
\includegraphics[width=.9\textwidth]{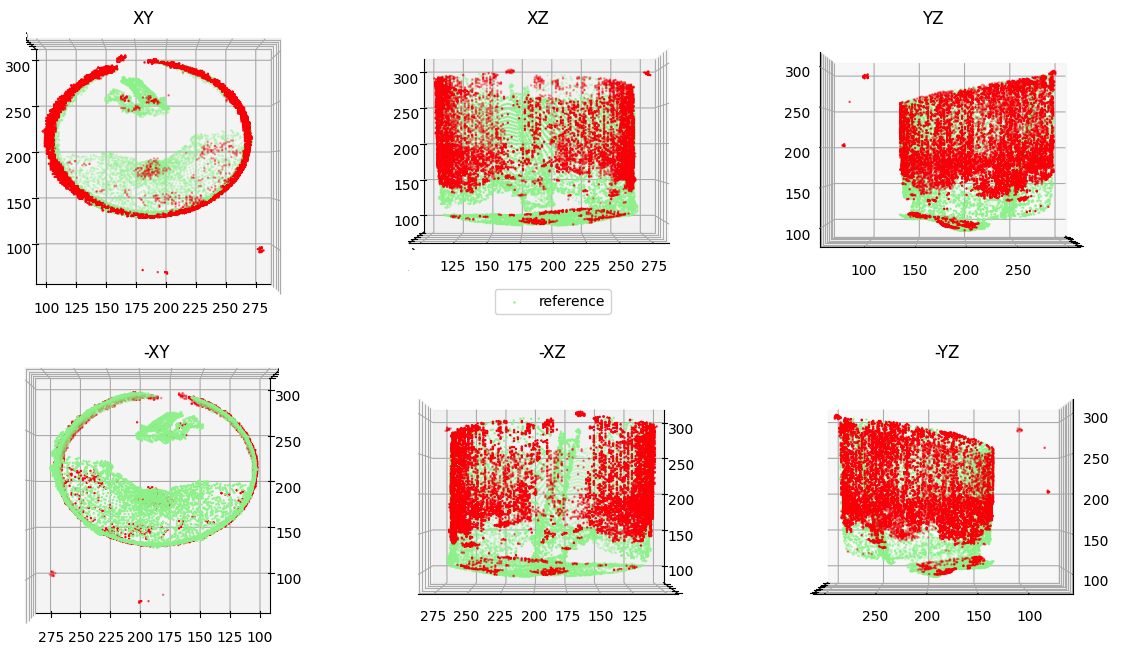}
\caption{Results from volume 5. Point cloud of the reference mesh label (in green) and the point cloud extracted by the proposed framework using a R2-Unet (in red).}
\label{fig:pc5}
\end{figure*}

\begin{figure*}[t]\centering
\includegraphics[width=.6\textwidth]{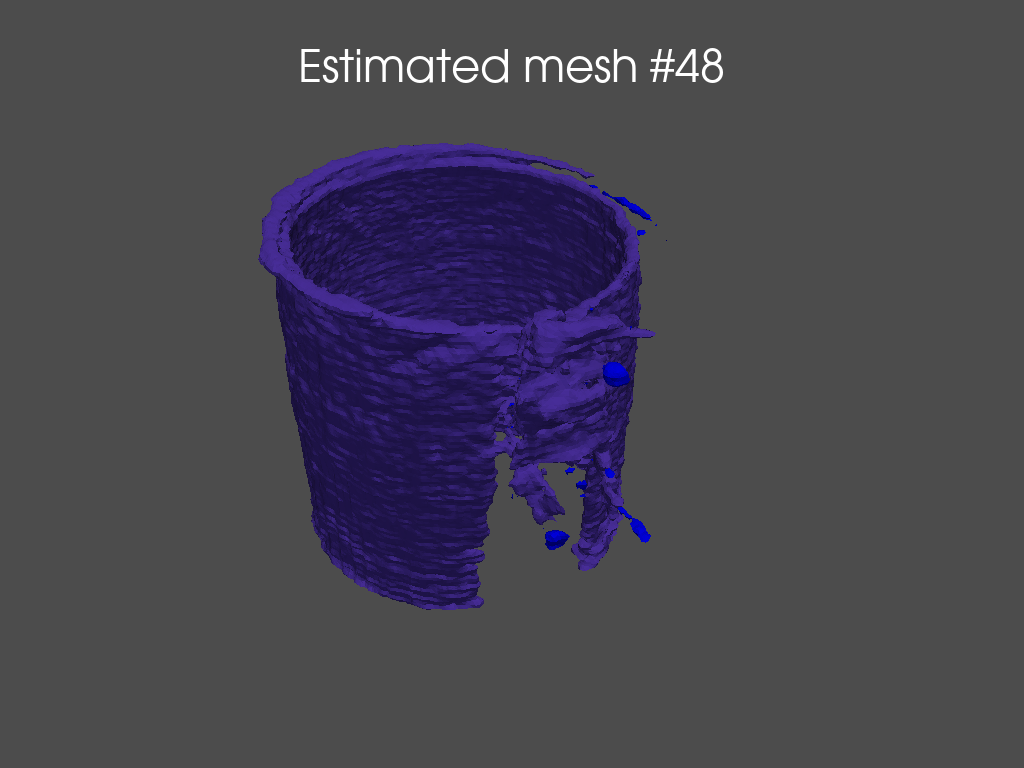}
\caption{Isosurface rendering of the points extracted for test volume 48 (whose reference label not provided to challenge participants).}
\label{fig:test48}
\end{figure*}

\begin{figure*}[t]\centering
\includegraphics[width=.9\textwidth]{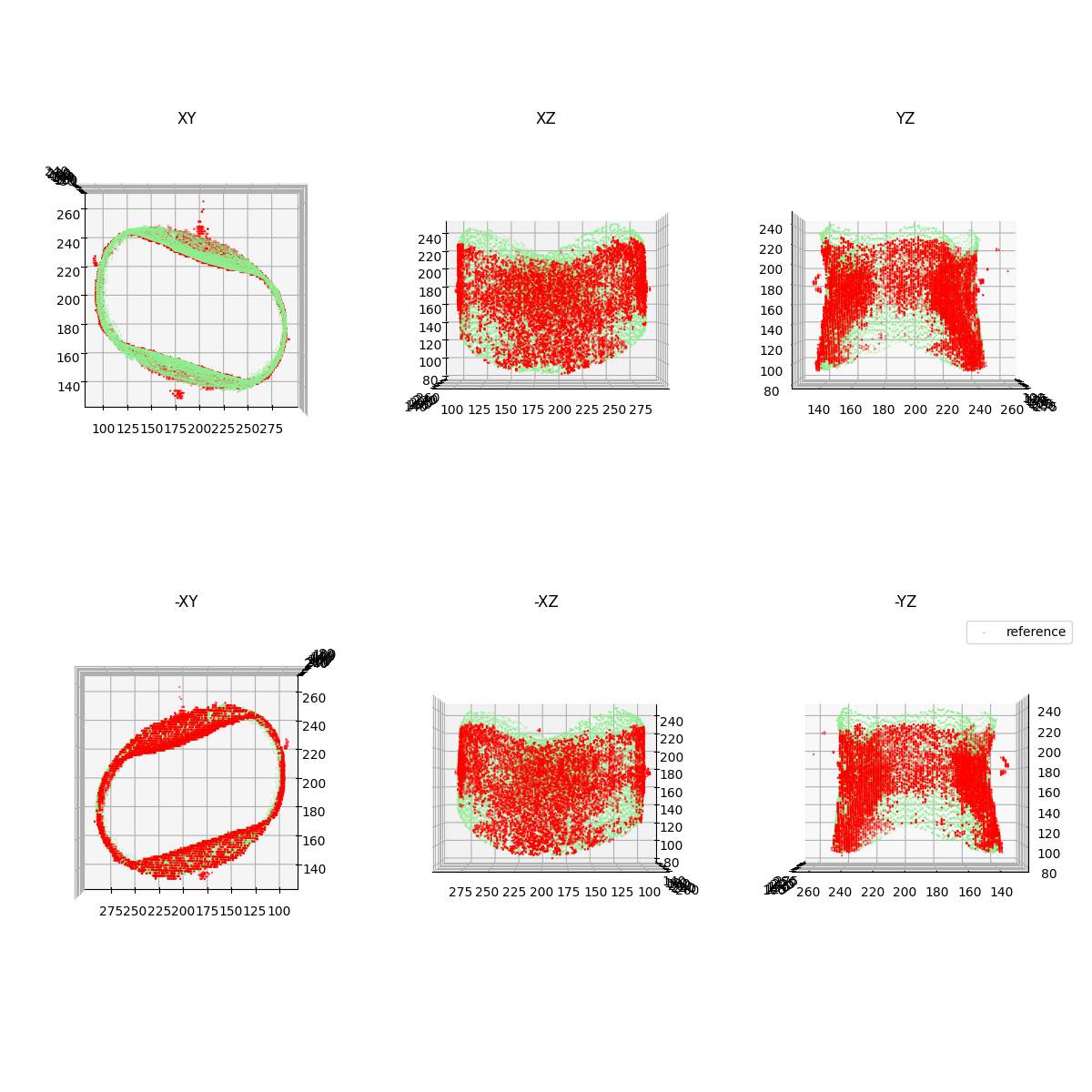}
\caption{Results for volume 1 by SE-Unet.}
\label{fig:pc1b}
\end{figure*}

\begin{figure*}[t]\centering
\includegraphics[width=.9\textwidth]{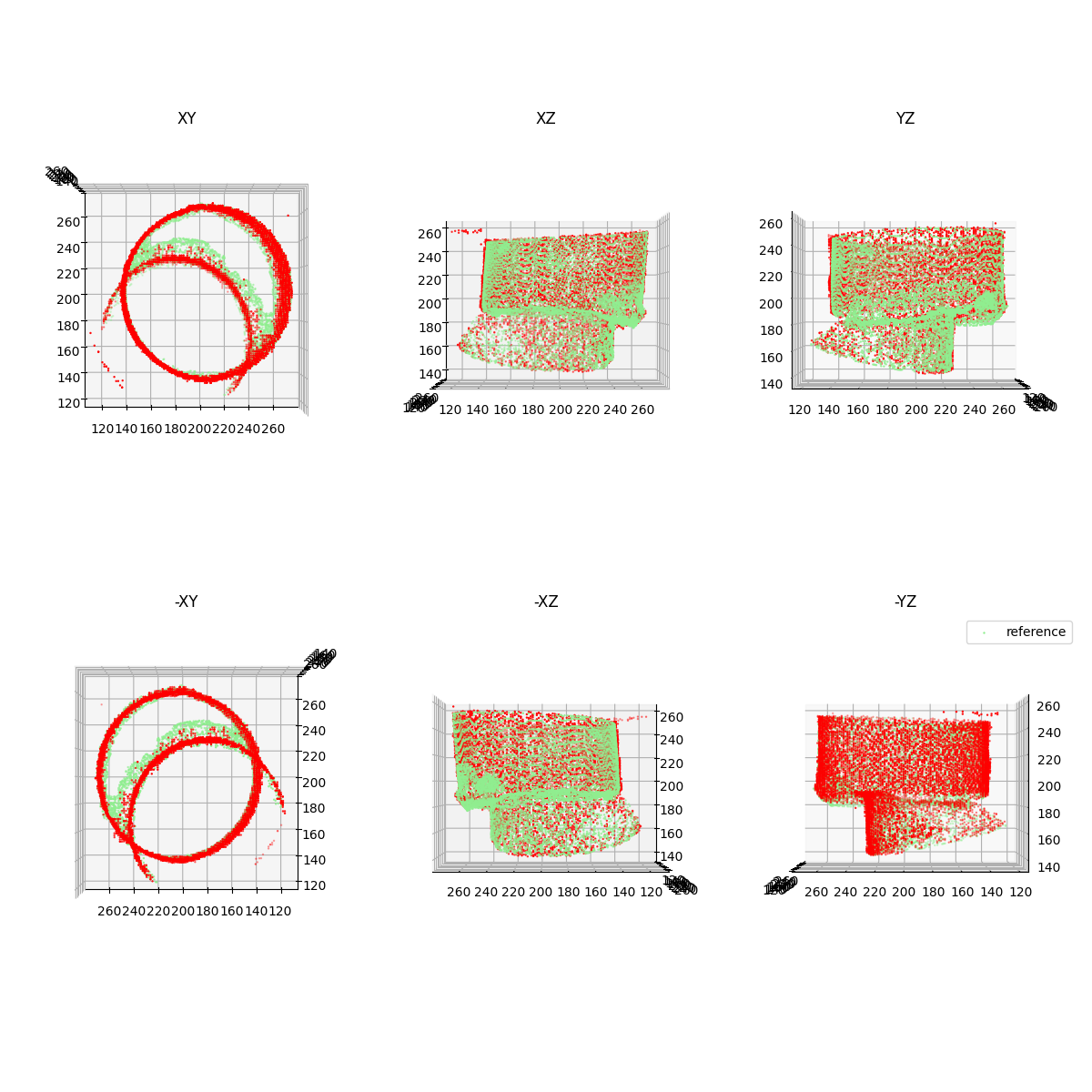}
\caption{Results for volume 2 using an SE-Unet.}
\label{fig:pc2b}
\end{figure*}

\begin{figure*}[t]\centering
\includegraphics[width=.9\textwidth]{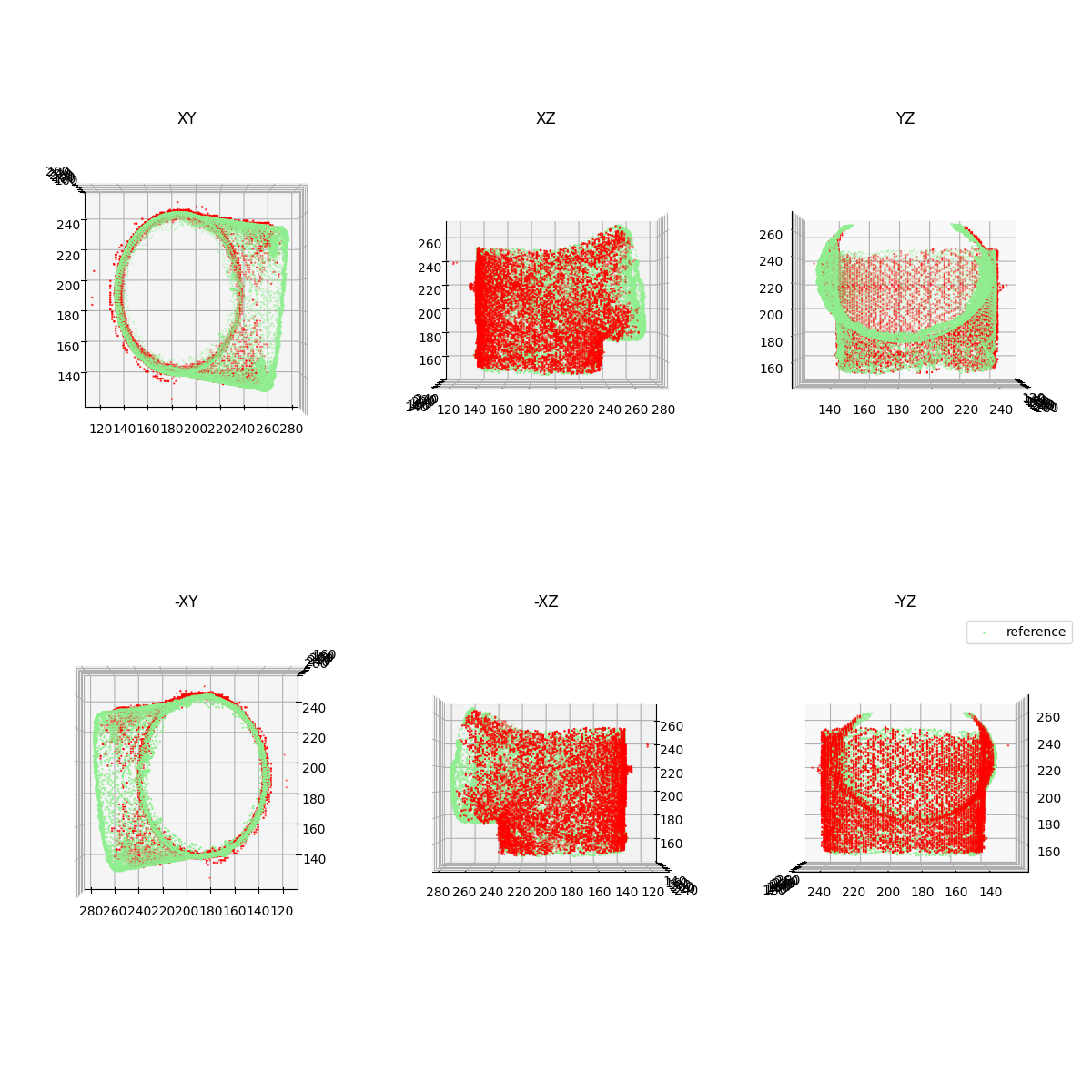}
\caption{Results for volume 3 using an SE-Unet.}
\label{fig:pc3b}
\end{figure*}

\begin{figure*}[t]\centering
\includegraphics[width=.9\textwidth]{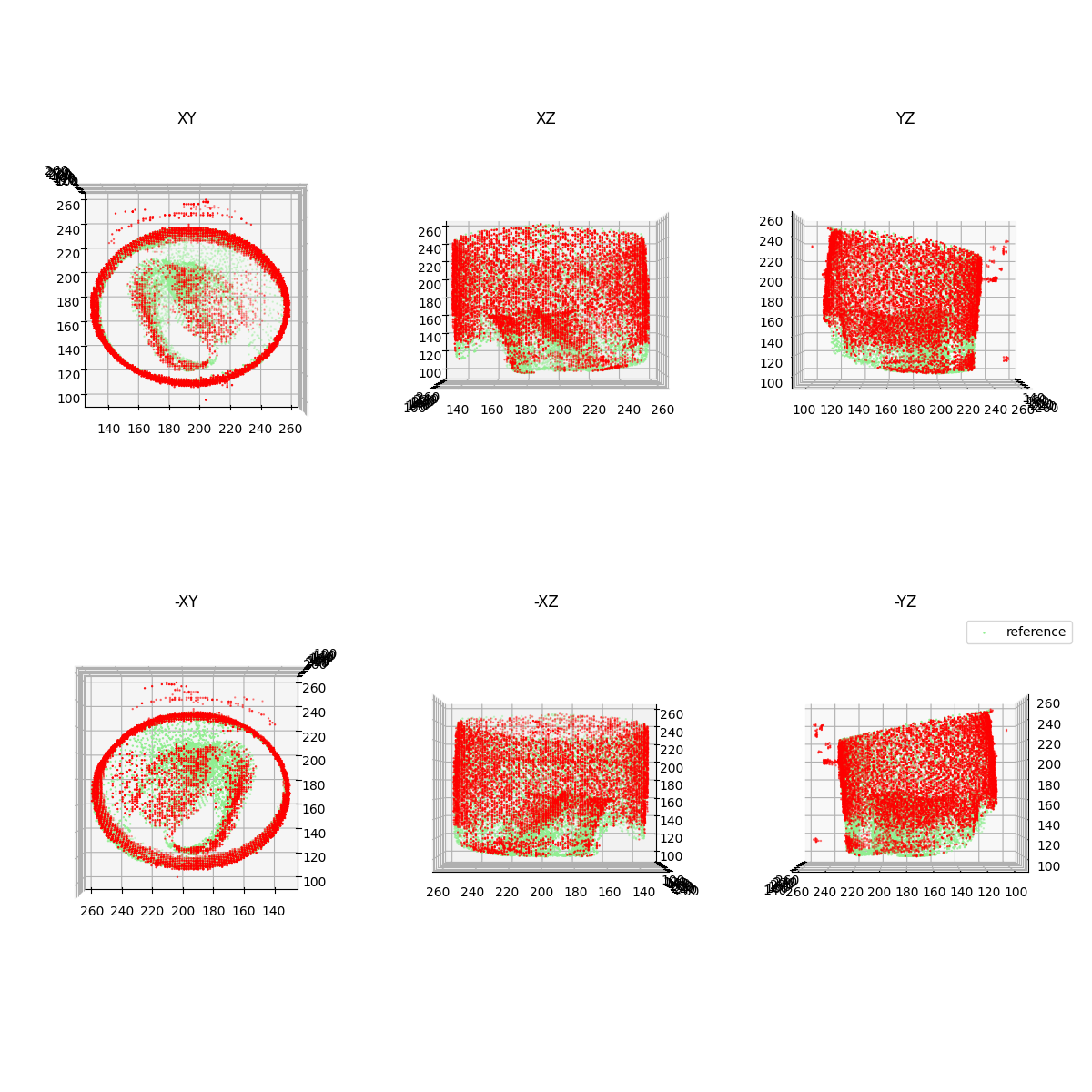}
\caption{Results for volume 4 using an SE-Unet.}
\label{fig:pc4b}
\end{figure*}

\begin{figure*}[t]\centering
\includegraphics[width=.9\textwidth]{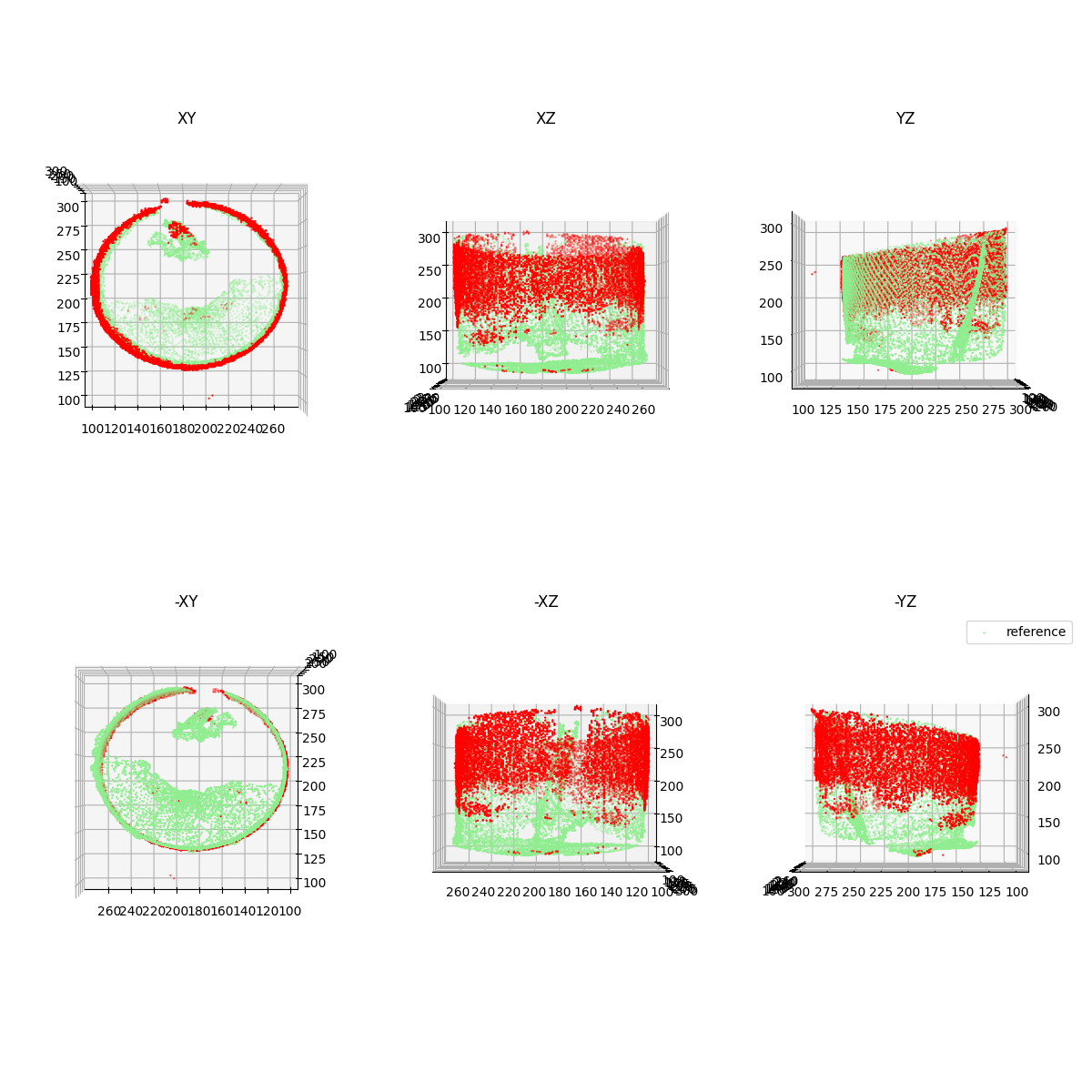}
\caption{Results for volume 5 using an SE-Unet.}
\label{fig:pc5b}
\end{figure*}

\newpage 
\section{Meta file}\label{sec:meta}

\begin{lrbox}{\codebox}
\begin{lstlisting}[language=python,label={lst:mhd},basicstyle=\small]
ObjectType = Image
NDims = 3
BinaryData = True
BinaryDataByteOrderMSB = False
CompressedData = False
TransformMatrix = 1.0 0.0 0.0 0.0 1.0 0.0 0.0 0.0 1.0
Offset = 0.0 0.0 0.0
CenterOfRotation = 0.0 0.0 0.0
AnatomicalOrientation = RAI
ElementSpacing = 0.49479 0.49479 0.3125
DimSize = 768 768 1280
ElementType = MET_USHORT
ElementDataFile = scan_001.raw
\end{lstlisting}
\end{lrbox}
\noindent
\colorbox{gray!10}{
\setlength{\fboxrule}{0ex}
    \fbox{%
      \begin{minipage}{\dimexpr\linewidth-0ex}
      \usebox{\codebox}    
      \end{minipage}%
    }
}

When the ultrasound data is read using a meta header file as demonstrated in Code listing \ref{lst:mhd}, users may need to update the values of the Anatomical Orientation tag.

\section{Training progress}\label{sec:progress}

\begin{figure*}[t]\centering
\includegraphics[width=.9\textwidth]{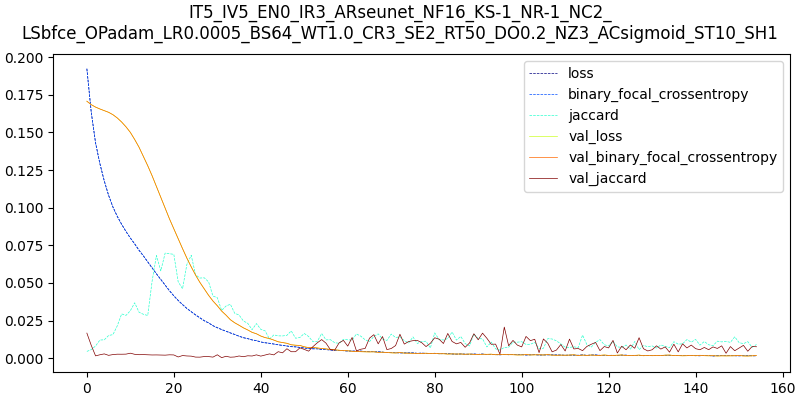}
\caption{Example progress plot of training a R2-Unet.}
\label{fig:progress}
\end{figure*}

\begin{figure*}[t]\centering
\includegraphics[width=.9\textwidth]{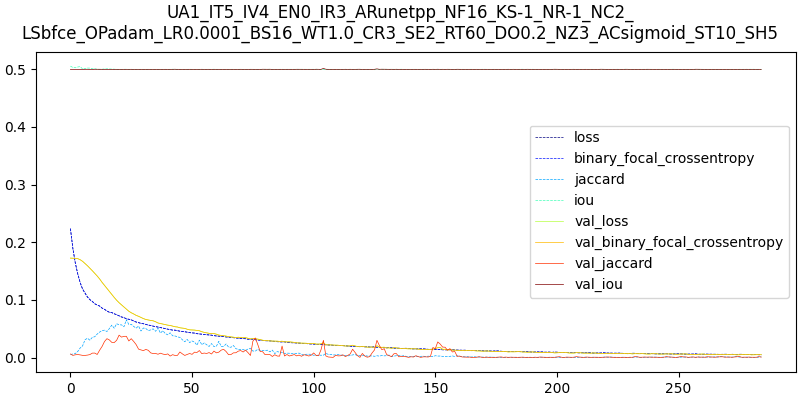}
\caption{Example progress plot of training an U-Net++.}
\label{fig:progress3}
\end{figure*}

\begin{figure*}[t]\centering
\includegraphics[width=.8\textwidth]{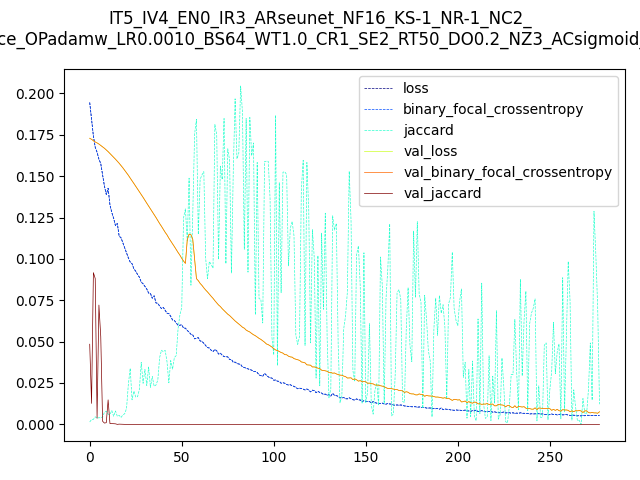}
\caption{Example progress plot of training an SE-Unet.}
\label{fig:progress2}
\end{figure*}

Figure \ref{fig:progress} and \ref{fig:progress2} show examples of training progress drawn from two randomly selected trials that involved a R2Unet and SE-Unet model.

\end{document}